# The Periodically Correlated Components of Measles in New York City using the Variable Band-pass Periodic Block Bootstrap

Anastazia Valachovic, Susan P. Opar, Edward Valachovic[1]

[1]Department of Epidemiology and Biostatistics, College of Integrated Health Sciences, University at Albany, State University of New York, Albany, New York, 12222

**Abstract**

Measles, a highly contagious, deadly virus, is at risk of losing its eradication status in the Unites States. Understanding the pattern, including seasonality, of measles could provide great benefit for forecasting, prevention, and public health preparedness as the virus re-emerges. The novel Variable Band-pass Periodic Block Bootstrap (VBPBB), which suppresses noise and interfering signals, is more efficient and statistically powerful to find periodic components compared to other existing methods. Using this method, we have found several significant periodic components in historical New York City measles cases which other methods were incapable of identifying. These findings give us a reusable method for understanding measles and other diseases and greater insight into what may occur should measles vaccine rates continue to fall.

## Introduction

Measles is a highly contagious virus with a very high basic reproduction number ($R_0$), the average number of additional cases caused by an infected individual, of approximately 12-18 (Anderson and May, 1982). Approximately 90% of non-immune individuals exposed to a case of measles will become infected (CDC, 2026). Measles is spread through respiratory droplets and can remain infective in a room for up to two hours (CDC, 2026). Symptoms of measles begin with fever; cough; runny nose; red, irritated eyes; subsequently followed by a red, spotty rash beginning on the face (CDC, 2026). As children are not able to be fully vaccinated until ages four to six years, children under the age of five are most vulnerable to more severe conditions caused by measles such as pneumonia, encephalitis, brain damage, and death (CDC, 2026). One to three out of every 1,000 children infected with measles will die from respiratory and/or neurologic complications (CDC, 2026). A rare, deadly complication of measles, subacute sclerosing panencephalitis (SSPE) can occur years later in children who appear to have recovered from measles. Non-immune pregnant and immune-compromised individuals are also at higher risk of measles complications. The measles vaccine is a live vaccine, and many immune-compromised individuals cannot receive live vaccines. Two doses of the measles vaccine are about 97% effective in preventing measles infection (CDC, 2026). There is no specific antiviral therapy approved by the Food and Drug Administration for treatment of measles; treatment is supportive and reflects the need for prevention. Furthermore, a systematic review found the average cost to address a single case of measles was approximately $43,000, highlighting not only the public health consequences of measles but also the economic burden (Sriudomporn and Patenaude, 2026).

The live measles vaccine was licensed in 1963. In 2000, measles was declared eliminated from the United States (CDC, 2026). Unfortunately, with declining vaccination rates and 2026 cases (as of August 2026: 2,777 cases) outnumbering 2025 cases (2,289), the United States is at risk of losing elimination status (CDC, 2026). The COVID-19 pandemic and vaccine hesitancy have led to increased numbers of measles outbreaks internationally (International Vaccine Access Center, 2025). Measles eradication requires exceptionally high levels of immunity in the population. Herd immunity must be maintained above 85 to 95 percent to halt broad transmission of measles (Katz and Hinman, 2000). As of June 30, 2026, 89% of New York City (NYC) children ages 24 to 35 months had received at least one measles vaccine via the MMR (measles, mumps, rubella vaccine) by their second birthday, a drop from 93% in 2024 (NYC Health, 2026). As of January 1, 2026, the New York statewide vaccine rate was 80.9% (NYS Department of Health, 2026).

Prior to the introduction of the measles vaccination, 90% of people were infected by 20 years of age, with an estimated 100 million cases and six million measles deaths yearly worldwide (Perry, 2015). It is estimated that the measles vaccine prevented approximately 57 million deaths worldwide from 2000 to 2022 (Minta, 2023). Unfortunately, millions of children were not vaccinated during the COVID-19 pandemic, resulting in an estimated 18% increase in measles cases and a 43% increase in estimated measles deaths in 2022 compared with 2021 worldwide (Minta, 2023).

As new outbreaks emerge and vaccination rates fluctuate it is important to understand the trends and patterns of measles rates across time. Many physical, environmental, biological, and human created events exhibit periodically correlated (PC) patterns when recorded as observations measured across time, also known as time series data. Time series definitions, notation, and examples can be found in Wei (1989). These PC component patterns can be related to influencing factors such as the 24-hour daylight cycle, the 7-day weekly work schedule, or the 12-month annual seasonal average temperature change. PC components, or those that repeat with a given period $p$ and whose reciprocal $1/p$ is called the frequency, exhibit strong correlation between observations that are $p$ time units apart. For instance, outdoor temperature at noon today is highly correlated with temperature at noon yesterday, a period of 24 hours ago, or an integer multiple of 24 hours. However, especially for those farther from the equator, the noon temperature also depends upon the season, so temperature at noon on July 1$^{st}$ one year, is more highly correlated with July 1st in other years, than temperature on any December 1st. Clearly, outdoor air temperature, with both daily PC and annual PC components, is an example of time series that are formed from multiple periodically correlated (MPC) components. Also, if a PC component defined by a given frequency $1/p$, called the fundamental frequency, has a repeating pattern that is not a sinusoid, it is often composed of changes occuring at integer $j$ multiples of the frequency $1/p$, called harmonics with frequency $j/p$. To better understand measles rates, it is necessary to investigate its PC components, harmonics, and the overall MPC pattern.

PC patterns are characterized by their period (or equivalent frequency) as well as their periodic mean, or the mean value at each point within the period. Bootstrapping can be a useful statistical analysis tool for investigating potential PC patterns. Bootstrapping is a non-parametric method of independently resampling from a data sample to estimate the sampling distribution of a statistic such as the mean and was first described by Efron (1979). Since time series data are ordered and often correlated, special adaptations of bootstrapping are necessary to preserve the correlation since independent random resamples of time series will not preserve the order of observations. The Non-overlapping Block Bootstrap (NBB) was introduced by Carlstein (1986), and a Moving Block Bootstrap (MBB) by Kunsch (1989). These helped replicate correlation between observations when bootstrapping time series data.

If a time series contains a PC component of period $p$, bootstrapping observations in blocks of most any length other than $p$, will not maintain the correlation of period $p$. The Seasonal Block Bootstrap (SBB) bootstraps blocks of the time series data sample in length $p$ to improve NBB and MBB for PC time series data and was proposed by Politis (2001). The SBB was improved by Dudek et al. (2014) in the General Seasonal Block Bootstrap (GSBB) and again in the Generalized Seasonal Tapered Block Bootstrap (GSTBB) by Dudek, Paparoditis, and Politis (2016).

These methods, collectively referred to here as Periodic Block Bootstraps (PBB), preserve the PC correlation but they bootstrap the original data sample which often contains noise, trends, and other PC components which can interfere with the PC component of interest. The Variable Band-pass Periodic Block Bootstrap (VBPBB) first band-pass filters the time series data sample, preserving the target frequency for one PC component, suppressing other interfering frequencies, and then bootstraps the reconstructed PC component. The VBPBB improves efficiency and statistical power to detect PC components over PBB and was introduced by Valachovic (2024). This method was extended to MPC components in the Variable Multiple Band-pass Periodic Block Bootstrap (VMBPBB) in Valachovic (2025).

While relatively new, the VBPBB and its MPC extension have successfully investigated and identified PC components undetectable by other PBB methods in diverse fields, demonstrating some clear advantages. They were first used to identify patterns in COVID deaths in Valachovic and Shishova (2025). They were used to investigate atmospheric particulate matter by Sun and Valachovic (2025), nitrogen dioxide pollution levels in DiMaio and Valachovic (2024), and electricity consumption in Yao and Valachovic (2024). They are also used in the fields of public health in Ahmad and Valachovic (2026) for COVID hospitalizations, and ischemic heart disease in Chen and Valachovic (2025). Ahmad et al. extended them for missing data (2025), Sun et al. (2025) investigated bias with them, and Yao et al. (2025) applied them to Bayesian analysis. For these reasons, VBPBB and its MPC extension are used here for the investigation into measles rates to help characterize the PC components and temporal patterns, with PBB used as a comparative analysis.

**Methods**

The New York City (NYC) Measles time series data spans 93 years, from 1891-1984, and is available from Hemple and Earn (2015). It relied heavily upon a NYC measles incidence rate dataset spanning 45 years, from 1928-1973, published by London and Yorke (1973). It expanded this dataset, added demographic information, included proportion of NYC population vaccinated, and increased temporal resolution from monthly to weekly observations. See Hemple and Earn (2015) for information concerning their process of data acquisition, compilation and quality-checking where they modeled NYC measles dynamics using the susceptible–infectious recovered (SIR) model. In this paper, weekly incidence rates (per 1 million) are normalized by computing weekly case counts divided by NYC population counts. Also, the NYC population records, recorded annually, are linearly interpolated to produce weekly population estimates. Finally, since data are bound below by 0 but have some very high values, similar to Hemple and Earn we transform the rate by taking the square root to control variance. The pre-vaccine NYC measles rate time series (1891–1963) is seen in Figure 1. For reference, the mean square root rate throughout the 1891-1963 pre-vaccine period is approximately seven, which translates to approximately 49 cases per 1 million population per week.

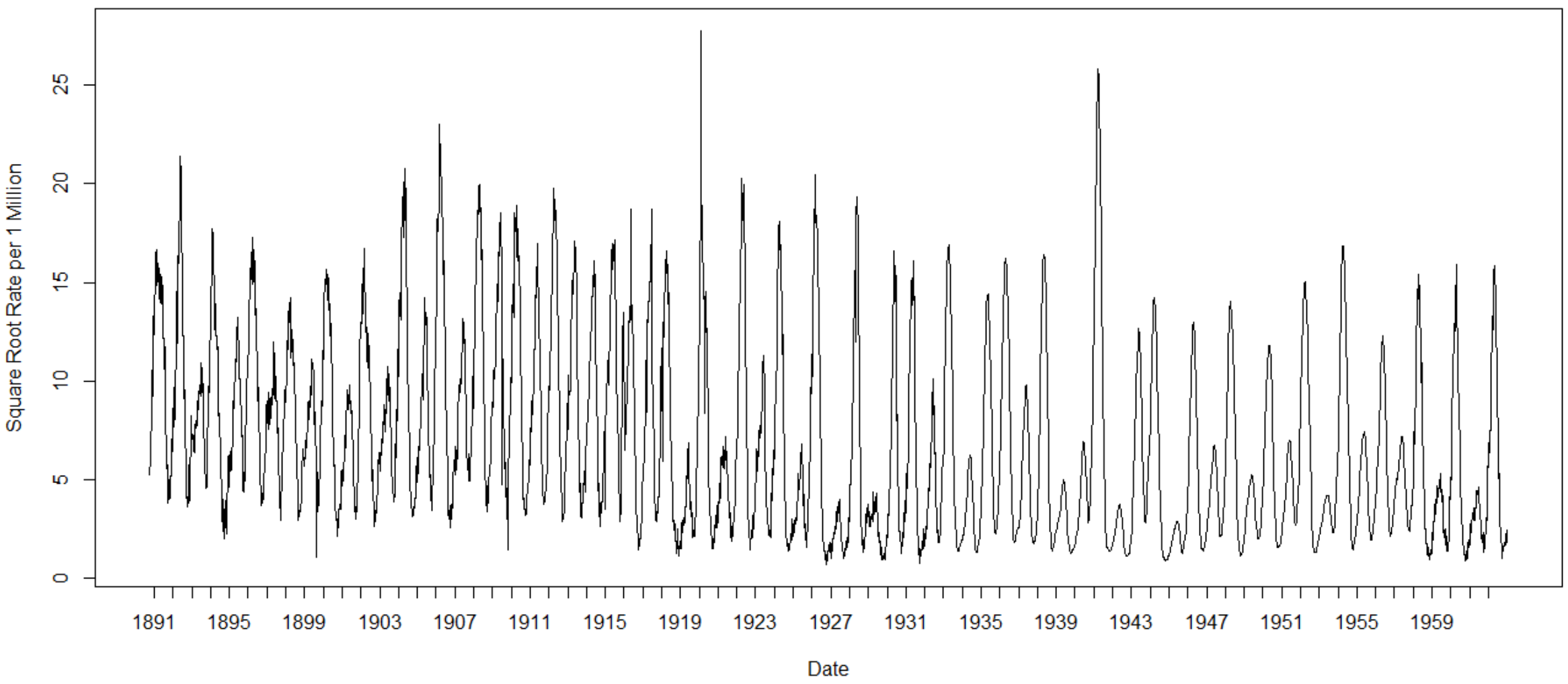

**Figure 1:** The pre-vaccine NYC measles rate time series.

Data importation, processing, and time series analysis were conducted in R version 4.1.1 (2013) statistical software using the KZFT function in the KZA package developed by Close and Zurbenko (2013). The Kolmogorov-Zurbenko Fourier Transform (KZFT) is an extension of the original Kolmogorov-Zurbenko (KZ) filter which is the iteration of a central simple central moving (SMA) average defined in Zurbenko (1986) with details in Yang and Zurbenko (2010). It is a filter with notation $KZ_{m,k}$ and two arguments, *m* and *k*. The argument $m$, a positive odd integer, is the length of the SMA filter window, and the argument *k*, a positive integer, is the number of iterations the SMA is applied. Therefore, the $SMA_m$ is a special case of the $KZ_{m,k}$ filter, when *k*=1. KZ filters are a type of low-pass filter that completely attenuates signals at the fundamental frequency 1/$m$ and well as its integer multiples or harmonics. It also strongly attenuates all signals of frequency 1/$m$ and higher while passing lower frequencies. The effect of a $KZ_{m,k}$ filter is smoothing the time series to variation occuring with period *m* or shorter.

An extension of the KZ filter, called the Kolmogorov-Zurbenko Fourier Transform (KZFT) is defined as an iteration of the Fourier Transform in Yang and Zurbenko (2010). The KZFT is a band-pass filter generalization of the KZ filter, with three functional arguments each with a clear interpretation. With notation $KZFT_{m,k,v}$ the arguments of the filter are *m*, the positive odd integer size or length of the filter window used in the filter calculations, *k*, the positive integer number of iterations of the Fourier transform, and *v*, the frequency at which the band-pass filter is centered. As such, the KZ filters are a special subset of KZFT filters when *v* = 0, in which case the band-pass is centered at a frequency of zero and it acts as a low-pass filter.

The Variable Band-pass Periodic Block Bootstrap (VBPBB) requires a band-pass filter step so it can periodic block bootstrap the reconstructed PC component with other components at different frequencies suppressed rather than the original time series composed of all components. However, it can be used with a variety of band-pass filters. Still, KZFT filters were used by Valachovic (2024) to design a new block bootstrapping method. Likewise, filtration and reconstruction of multiple PC components with specific KZFT filters were instrumental in the design of a first of its kind method for resampling multiple periodically correlated (MPC) components in time series called the Variable Multiple Band-pass Periodic Block Bootstrap (VMBPBB) (Valachovic, 2025). Here we also use the KZFT for band-pass filters used in VBPBB and VMBPBB. The KZFT filters parameters are set according to Valachovic (2024) for each tested PC component, where it is noted that PBB is a special, trivial, case of VBPBB when the band-pass filter is so wide all frequencies pass and none are suppressed. However, PBB bootstraps unrelated frequencies, increasing variation and often underestimating the true nominal percent confidence of the confidence interval (CI), thus creating a CI size that is too large for its stated confidence. The arguments of the KZFT filter are used to narrow the band-pass filter region, lowering the pass band, suppressing interfering frequencies prior to bootstrapping. So VBPBB can decrease the nominal percent CI along with the CI size, providing the means to make the CI produced equal to the stated nominal percent. This is done by the appropriate choice of KZFT arguments which control the band-pass filter. Following

Valachovic (2024) for the KZFT argument selection for each VBPBB application, argument *v* is set equal to the frequency of the PC component being tested. Valachovic (2024) notes both *k* and *m* change the pass band width of the band-pass filter, however increasing the number of iterations *k* generally demands more available data to apply the iterated filter completely as compared to increasing *m*. Consequently, argument *k* is set to one. Finally, argument *m* is chosen according to the particular combination of the period of the PC component of interest, the number of cycles observed in the time series, and the relative signal strength compared to all other components, or the signal-to-noise ratio. Knowing this information, Valachovic (2024), suggests the optimal *m* choice to produce a KZFT band-pass filter of particular characteristics to produce a CI that is 95% with VBPBB. The result of VBPBB is a 95% confidence interval band, or a set of 95% CI with one at each time point in the PC component period. A significant PC component is one having a 95% CI band that excludes the possibility of no variation, or a flat line fitting through the CI band, at the selected frequency. If the 95% CI band for a PC component excludes the possibility of no variation, or a flat line, then it is significant at the 95% level and a 95% CI for the amplitude is the highest 95% CI in the CI band.

PC components are selected for VBPBB testing for significance for several reasons. The first source is from theory and prior research. For instance, many natural processes operate on an annual cycle (52 weeks). Likewise, some processes are influenced by human made factors, for example, a two or four-week pay period cycle. The next sources are those frequencies (or periods) suggested by the periodogram by high peaks. A periodogram is the estimate of the spectral density or the energy at each frequency contained within a given time series. A portion of the periodogram for pre-vaccine NYC measles rates is given in Figure 2.

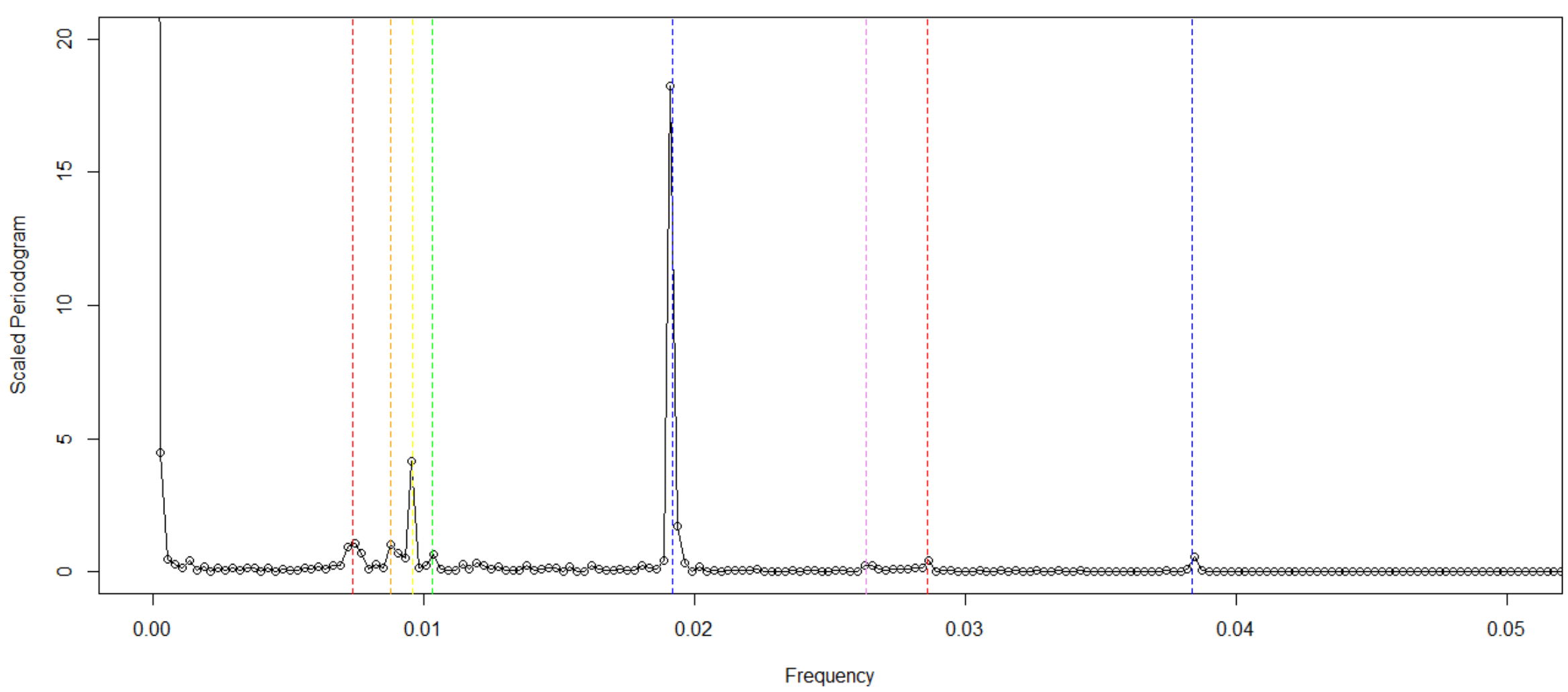

**Figure 2:** The pre-vaccine NYC measles rate time series periodogram. Frequencies are indicated from left to right for 136-weeks in red, 114-weeks in orange, 2-year (104-weeks) in yellow, 97-weeks in green, annual (52-weeks) in blue, 38-weeks in violet, 35-weeks in red, and annual second harmonic (26-weeks) in blue.

With the exclusion of a peak at zero corresponding to a trend in the data, starting with the largest, a subset of the notable PC peaks are at frequency 1/52, 1/104, and 1/26, corresponding to annual, 2-year, and half-year periods respectively. The last source of potential PC components are harmonics of the aforementioned potential fundamental frequencies. Using theory, the periodogram, and harmonics, this study compiles a list of potential PC frequencies (periods) that is not exhaustive, but for this research includes 136-weeks, 114-weeks, 2-year (104-weeks), 97-weeks, annual (52-weeks), 38-weeks, 35-weeks, annual second harmonic (26-weeks), annual third harmonic (~17-weeks), annual fourth harmonic (~13-weeks), annual fifth harmonic (~10-weeks), annual sixth harmonic (~9-weeks), 4-weeks, and 2-weeks.

**Results**

The following Table 1 summarizes the results of the pre-vaccine NYC measles data, showing tested and significant PC components. The first column represents the period in weeks (the reciprocal of frequency) for the tested potential periodic components. For each PC tested, the interval in the table is the 95% CI of the highest time point throughout the period. If this interval does not contain zero, then this indicates the PC has an amplitude greater than zero, suggesting statistical significance. Here, statistical significance is indicated by bolded text and an asterisk. Also provided are the signal to noise ratio of the PC component and the KZFT argument m used in the VBPBB process. VBPBB found nine out of the fourteen tested potential PC components to be statistically significant. The table additionally notes the 104-week PC component was the only component found to be significant by the PBB method used for comparison. All other PC components were insignificant using the PBB method.

**Table 1:** Tested periods of potential PC components in pre-vaccine NYC measles rate with 95% CI for the amplitude and arguments used in the VBPBB analysis.

| Period in Weeks | VBPBB Pre-Vaccine Square Root Rate 95% CI | Signal to Noise Ratio and KZFT m argument used in VBPBB |
|---|---|---|
| 136-weeks | **(0.38,2.55)***  | S/N=0.04 M=1501 |
| 114-weeks | **(0.44,1.85)*** | S/N=0.04 M=1501 |
| 104-weeks (2-years) | **(1.32,3.79)** * † | S/N=0.07 M=1501 |
| 97-weeks | **(0.10,1.45)*** | S/N=0.04 M=1501 |
| 52-weeks (annual) | **(2.82,5.05)*** | S/N=0.59 M=1451 |
| 38-weeks | **(0.25,1.16)*** | S/N=0.006 M=1451 |
| 35-weeks | **(0.07,1.17)*** | S/N=0.007 M=1501 |
| 26-weeks (annual 2nd harmonic) | **(0.35,1.06)*** | S/N=0.016 M=1451 |
| 17-weeks (annual 3rd harmonic) | (-0.09,0.097) | S/N=0.00001 M=1451 |
| 13-weeks (annual 4th harmonic) | (-0.05,0.19) | S/N=0.00008 M=1501 |
| 10-weeks (annual 5th harmonic) | (-0.00003,0.11) | S/N=0.000008 M=1701 |
| 9-weeks (annual 6th harmonic) | **(0.01,0.06)*** | S/N=1.6e-05 M=1701 |
| 4-weeks | (-0.02,0.01) | S/N=3.9e8-08 M=1701 |
| 2-weeks | (-0.02,0.05) | S/N=3.7e-07 M=1701 |

**Note:** * indicates bolded intervals that are significant at the 95% confidence level using VBPBB. † indicates significant at the 95% confidence level using PBB.

The next set of results include descriptions and figures for examples of significant PC components, their temporal patterns, and both the 95% CI bands produced by VBPBB and the less statistically powerful PBB methods.

The 136-week component was found to be significant using VBPBB and a KZFT argument of m=1501. It had an estimated signal to noise ratio of 0.04, and provided an estimated 95% CI for the amplitude of the component from (0.38, 2.55). For comparison, PBB could not identify this component as significant. Figure 3 shows several cycles of the 136-week component, the 95% CI band from VBPBB in blue (which excludes a flat line of insignificance) and the KZFT band-pass filtered PC component in green that was used for bootstrapping with VBPBB. Figure 3 also shows the 95% CI band from PBB in red and the original data in yellow used for bootstrapping with PBB.

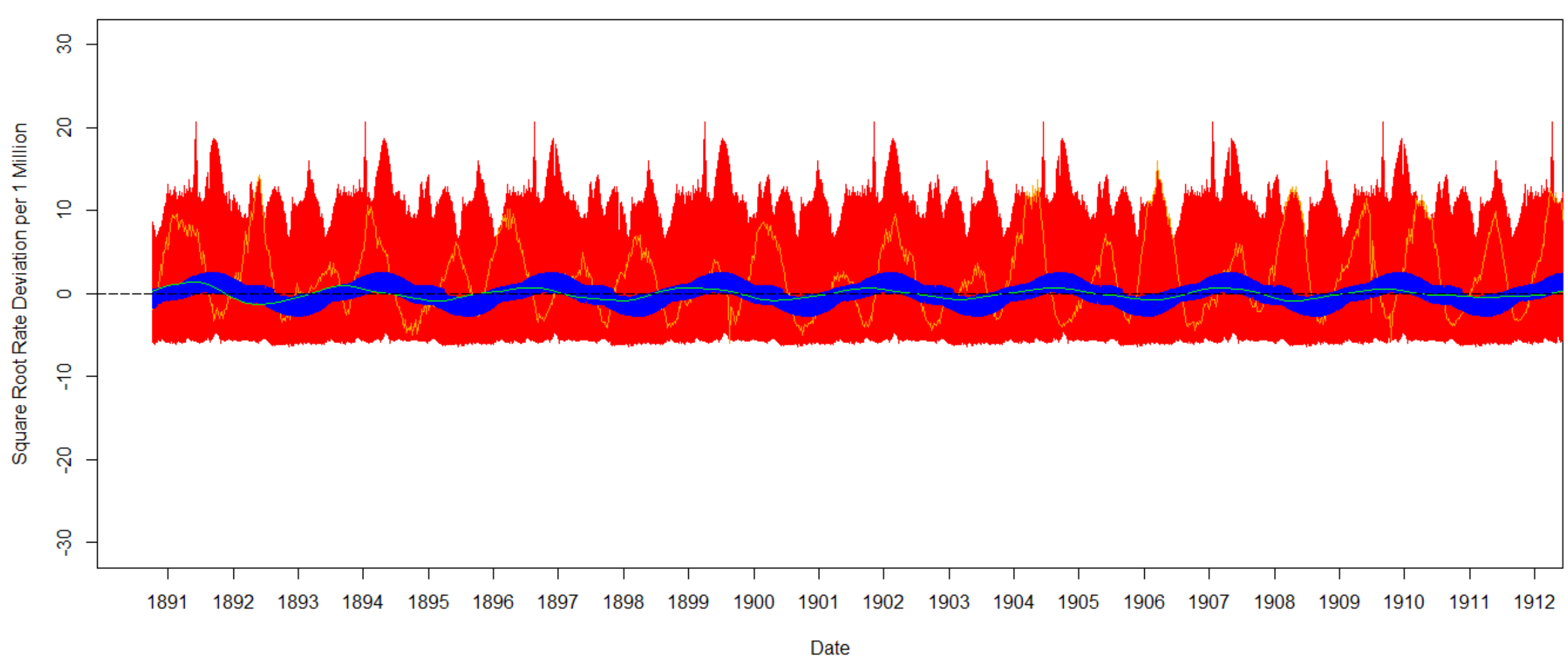

**Figure 3:** Several cycles of the 136-week component of pre-vaccine NYC measles rate. The 95% CI band from VBPBB is in blue and the KZFT band-pass filtered PC component is in green. The 95% CI band from PBB is in red and the original data is in yellow.

The 114-week component was found to be significant using VBPBB and a KZFT argument of m=1501. It had an estimated signal to noise ratio of 0.04 and provided an estimated 95% CI for the amplitude of the component from (0.44, 1.85). Again, for comparison, PBB could not identify this component as significant. Figure 4 shows several cycles of the 114-week component, the 95% CI band from VBPBB in blue (which excludes a flat line of insignificance) and the KZFT band-pass filtered PC component in green that was used for bootstrapping with VBPBB. The figure also shows the 95% CI band from PBB in red and the original data in yellow used for bootstrapping with PBB.

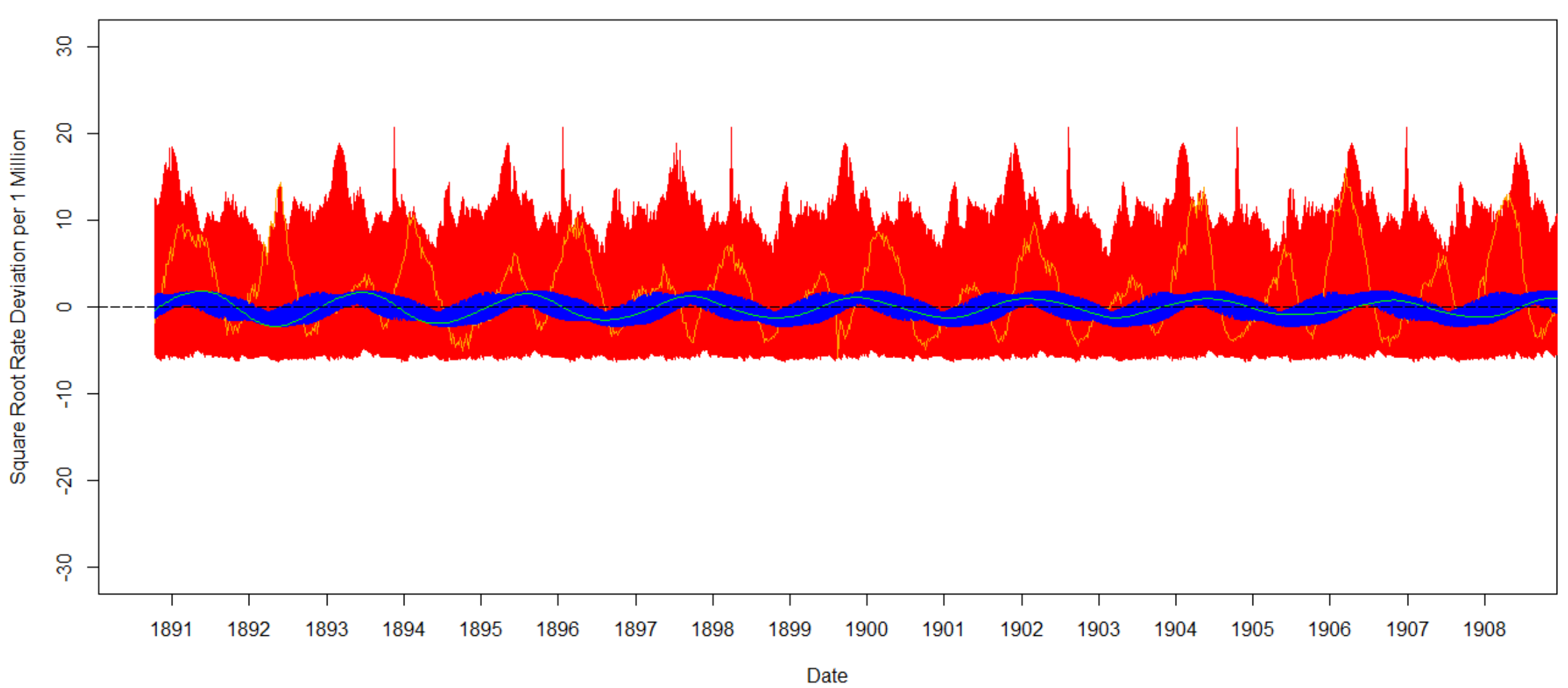

**Figure 4:** Several cycles of the 114-week component of pre-vaccine NYC measles rate. The 95% CI band from VBPBB is in blue and the KZFT band-pass filtered PC component is in green. The 95% CI band from PBB is in red and the original data is in yellow.

The 104-week component was found to be significant using VBPBB and a KZFT argument of m=1501. It had an estimated signal to noise ratio of 0.07 and provided an estimated 95% CI for the amplitude of the component from (1.32, 3.79). PBB was able to identify this component as significant. Figure 5 shows several cycles of the 104-week component, the 95% CI band from VBPBB in blue (which excludes a flat line of insignificance) and the KZFT band-pass filtered PC component in green that was used for bootstrapping with VBPBB. The figure also shows the 95% CI band from PBB in red and the original data in yellow used for bootstrapping with PBB.

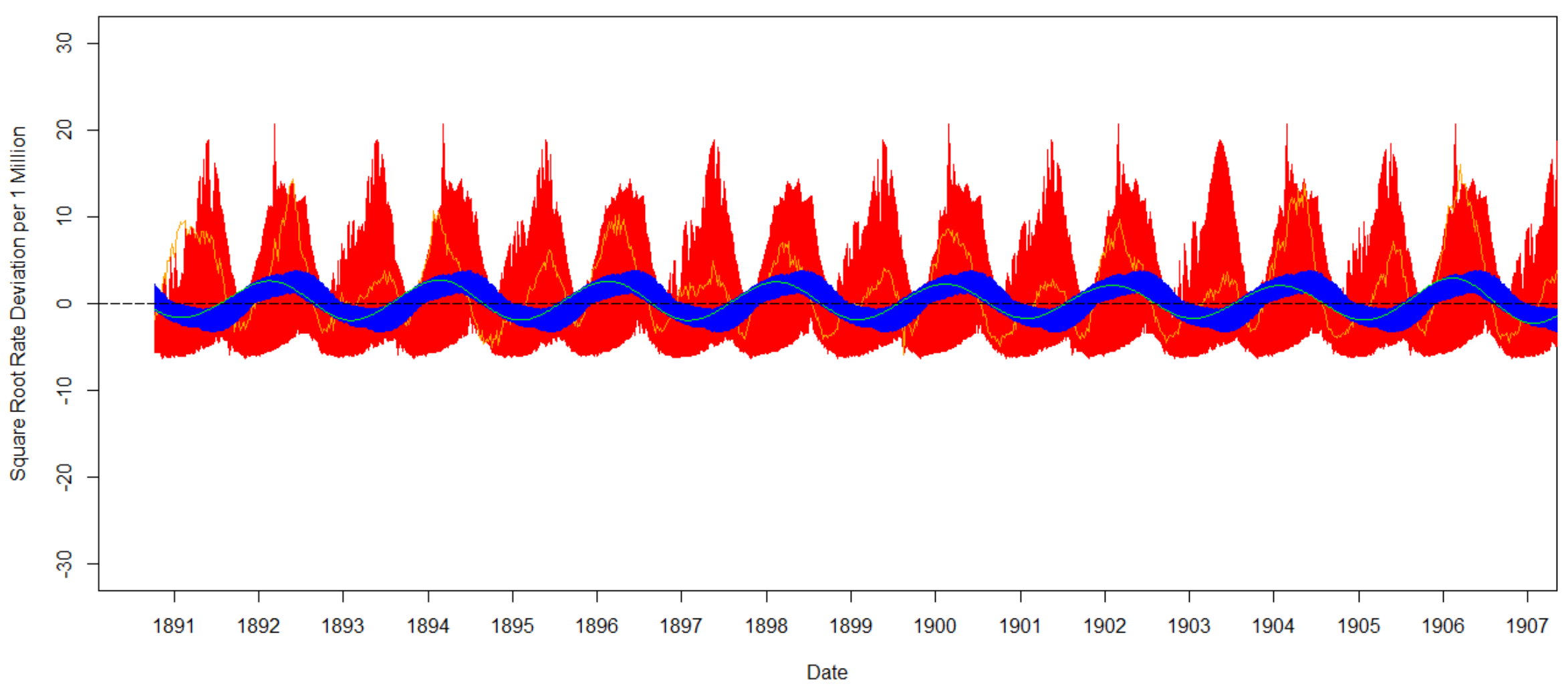

**Figure 5:** Several cycles of the 104-week component of pre-vaccine NYC measles rate. The 95% CI band from VBPBB is in blue and the KZFT band-pass filtered PC component is in green. The 95% CI band from PBB is in red and the original data is in yellow.

The 97-week component was found to be significant using VBPBB and a KZFT argument of m=1501. It had an estimated signal to noise ratio of 0.04 and provided an estimated 95% CI for the amplitude of the component from (0.10, 1.45). Again, for comparison, PBB could not identify this component as significant. Figure 6 shows several cycles of the 97-week component, the 95% CI band from VBPBB in blue (which excludes a flat line of insignificance) and the KZFT band-pass filtered PC component in green that was used for bootstrapping with VBPBB. The figure also shows the 95% CI band from PBB in red and the original data in yellow used for bootstrapping with PBB.

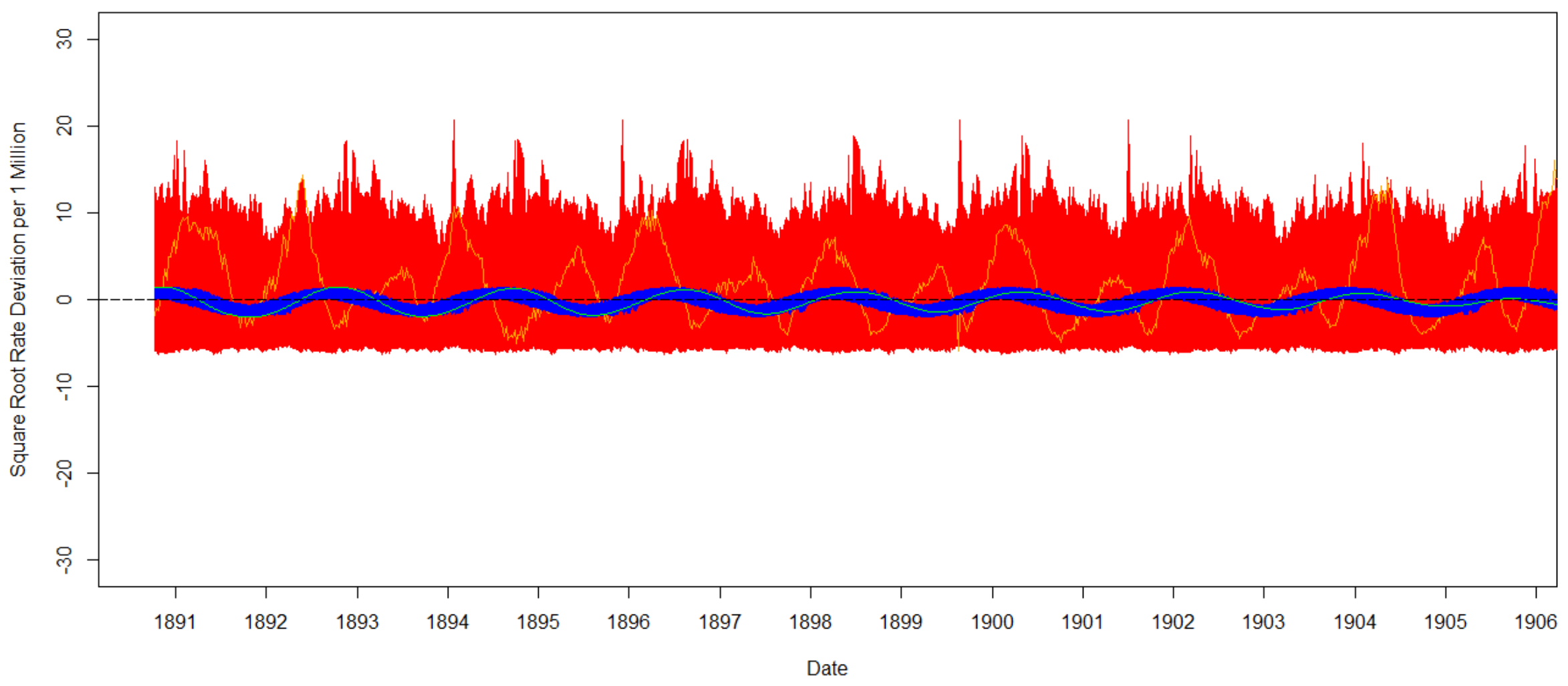

**Figure 6:** Several cycles of the 97-week component of pre-vaccine NYC measles rate. The 95% CI band from VBPBB is in blue and the KZFT band-pass filtered PC component is in green. The 95% CI band from PBB is in red and the original data is in yellow.

The 52-week component was found to be significant using VBPBB and a KZFT argument of m=1451. It had an estimated signal to noise ratio of 0.59 and provided an estimated 95% CI for the amplitude of the component from (2.82, 5.05). Again, for comparison, PBB could not identify this component as significant. Figure 7 shows several cycles of the 52-week component, the 95% CI band from VBPBB in blue (which excludes a flat line of insignificance) and the KZFT band-pass filtered PC component in green that was used for bootstrapping with VBPBB. The figure also shows the 95% CI band from PBB in red and the original data in yellow used for bootstrapping with PBB.

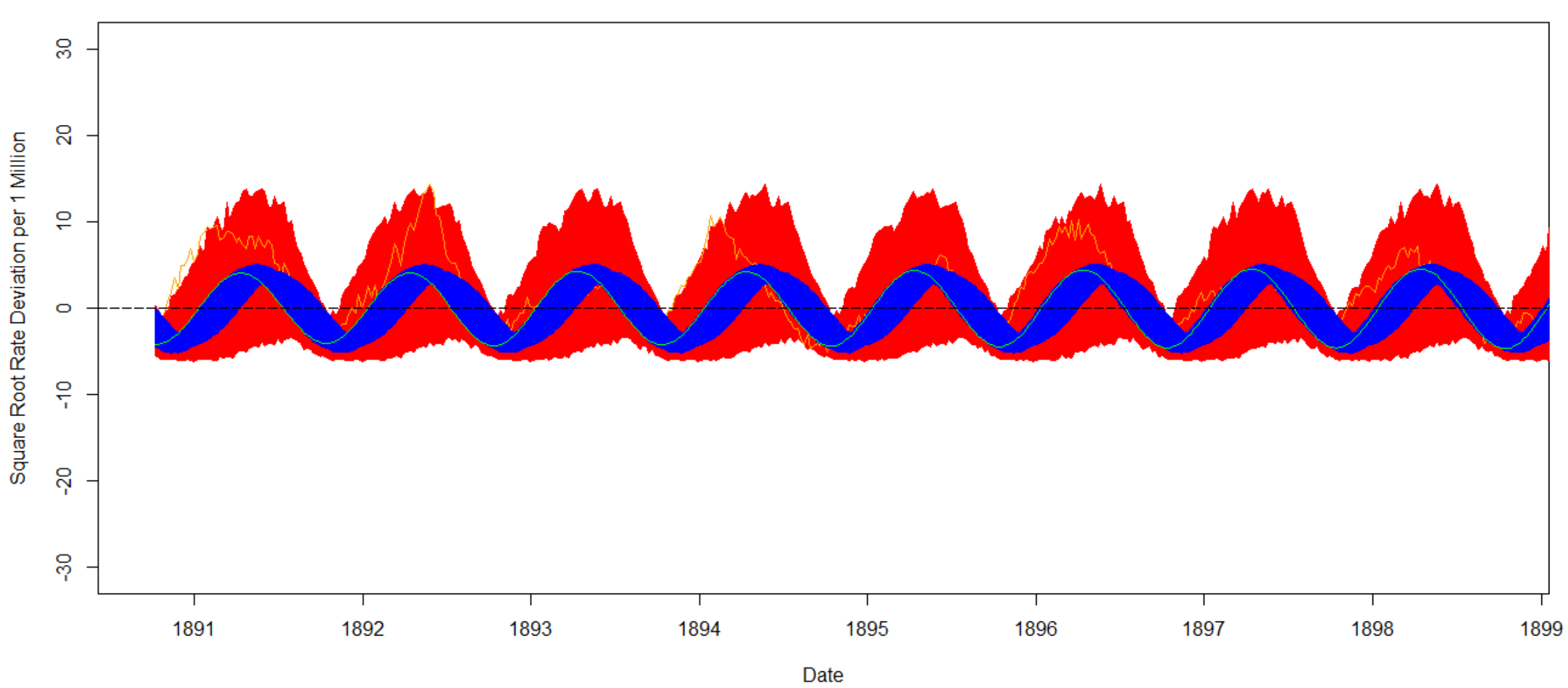

**Figure 7:** Several cycles of the 52-week component of pre-vaccine NYC measles rate. The 95% CI band from VBPBB is in blue and the KZFT band-pass filtered PC component is in green. The 95% CI band from PBB is in red and the original data is in yellow.

The 26-week component was found to be significant using VBPBB and a KZFT argument of m=1451. It had an estimated signal to noise ratio of 0.016 and provided an estimated 95% CI for the amplitude of the component from (0.35, 1.06). Again, for comparison, PBB could not identify this component as significant. Figure 8 shows several cycles of the 26-week component, the 95% CI band from VBPBB in blue (which excludes a flat line of insignificance) and the KZFT band-pass filtered PC component in green that was used for bootstrapping with VBPBB. The figure also shows the 95% CI band from PBB in red and the original data in yellow used for bootstrapping with PBB.

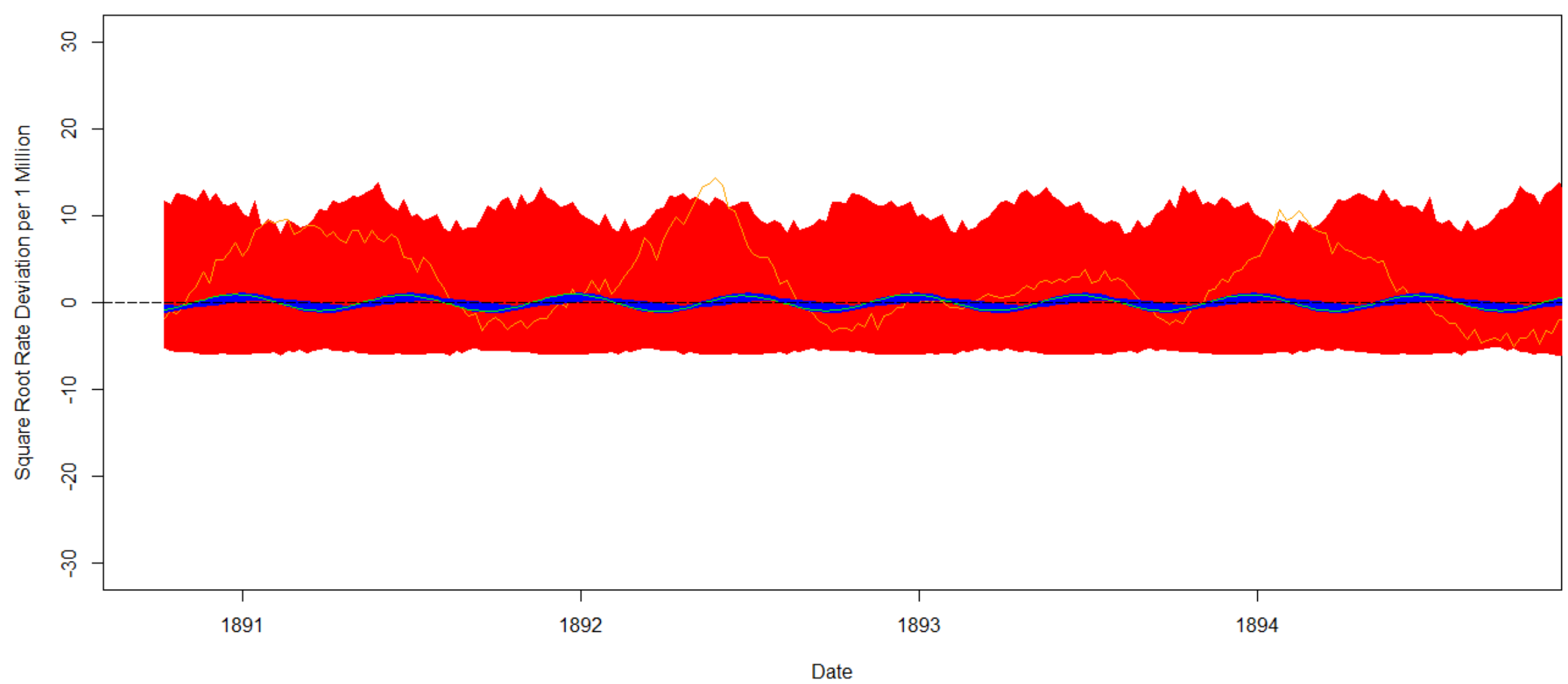

**Figure 8:** Several cycles of the 26-week component of pre-vaccine NYC measles rate. The 95% CI band from VBPBB is in blue and the KZFT band-pass filtered PC component is in green. The 95% CI band from PBB is in red and the original data is in yellow.

The following Figure 9 illustrates the nine significant PC components found from VBPBB combined to give a 95% CI band in blue for the MPC components in the pre-vaccine NYC measles data. The 95% CI band in red from PBB is only for comparative purposes and should be noted it is the CI band for only one PC component (the 52-week component) and not all nine significant components found through VBPBB. This illustrates the 95% CI band for all nine significant PC components from VBPPB is generally smaller than the 95% CI band for just one component from PBB, showing the strength of VBPBB for both PC and MPC analysis.

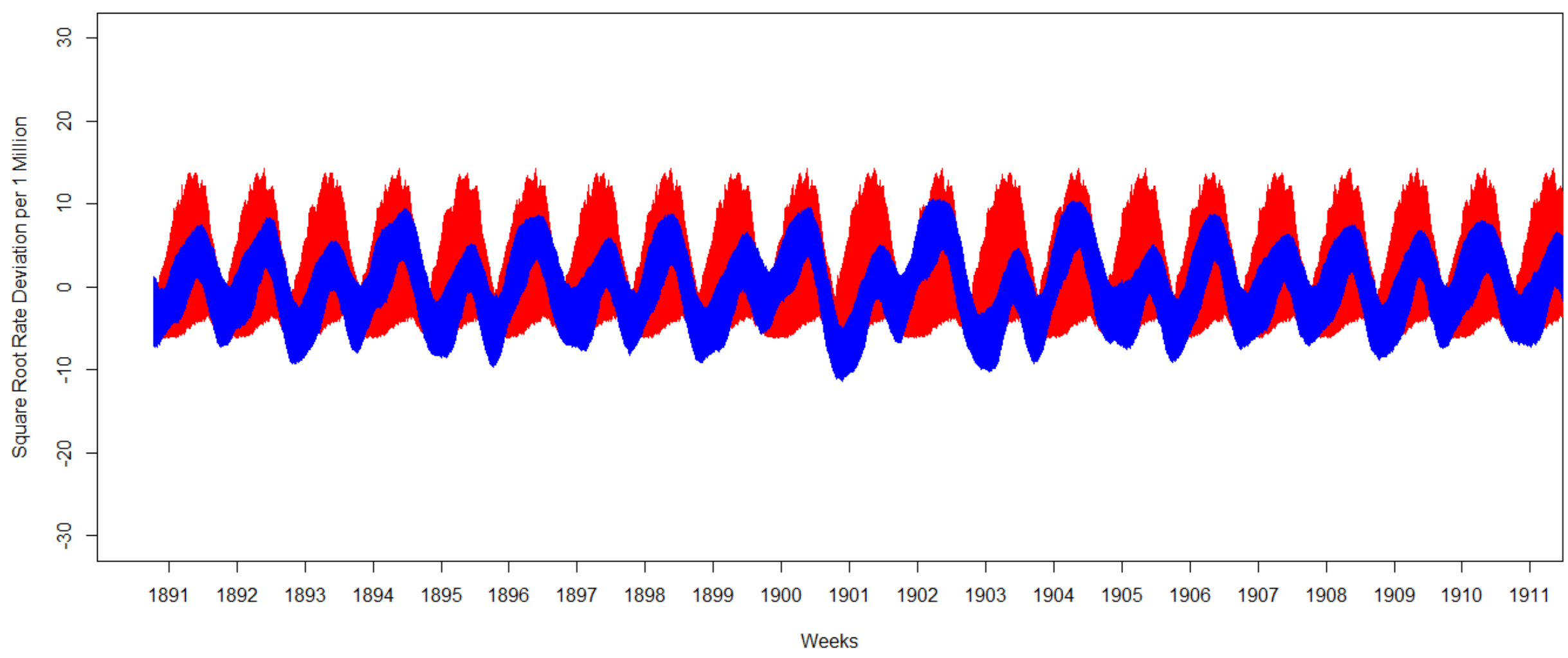

**Figure 9:** Several cycles of all significant PC components of the pre-vaccine NYC measles rate. The 95% CI band from VBPBB is in blue. The 95% CI band from PBB for the 52-week component only is in red.

This next set of results pertains to the period of NYC measles rates after the introduction of the measles vaccine. The NYC measles rate post-vaccine introduction time series (1963 – 1984), after a square root transformation, is seen in Figure 10. For reference, the mean square root rate throughout the 1963-1984 post-vaccine introduction period is approximately 1.79, which translates to approximately 3.20 cases per 1 million population per week.

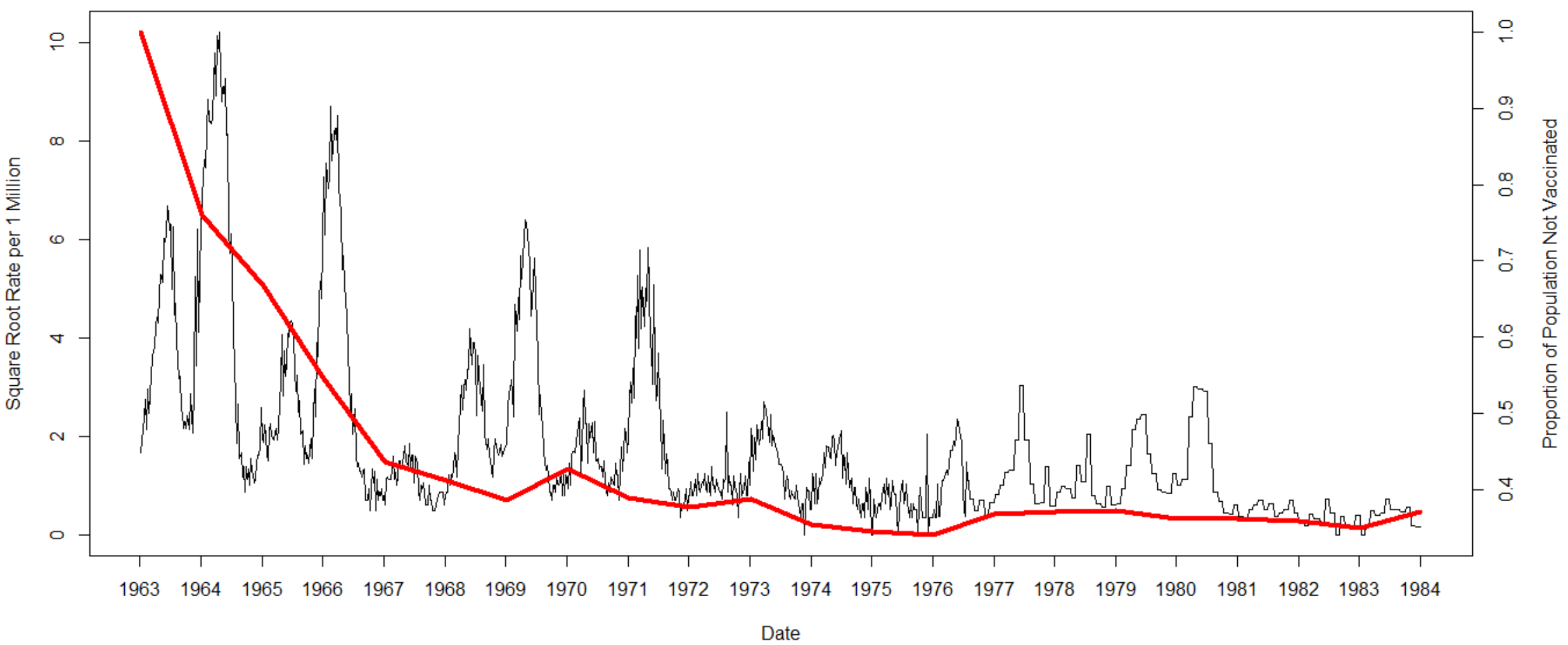

**Figure 10:** The post-vaccine introduction NYC measles rate time series. The square root measles rate per 1 million is in black and in red is the proportion of the NYC population not vaccinated.

A portion of the periodogram for the post-vaccine introduction NYC measles rates is given in Figure 11. The notable PC peaks post-vaccine introduction appear different than those pre-vaccine introduction. In the periodogram, some peak frequencies remain such as at frequency 1/52, however, the size of this peak is greatly reduced. Others appear to have disappeared. To compare the pre- and post-vaccine introduction time spans, the same candidate PC components are retested for significance using VBPBB and PBB.

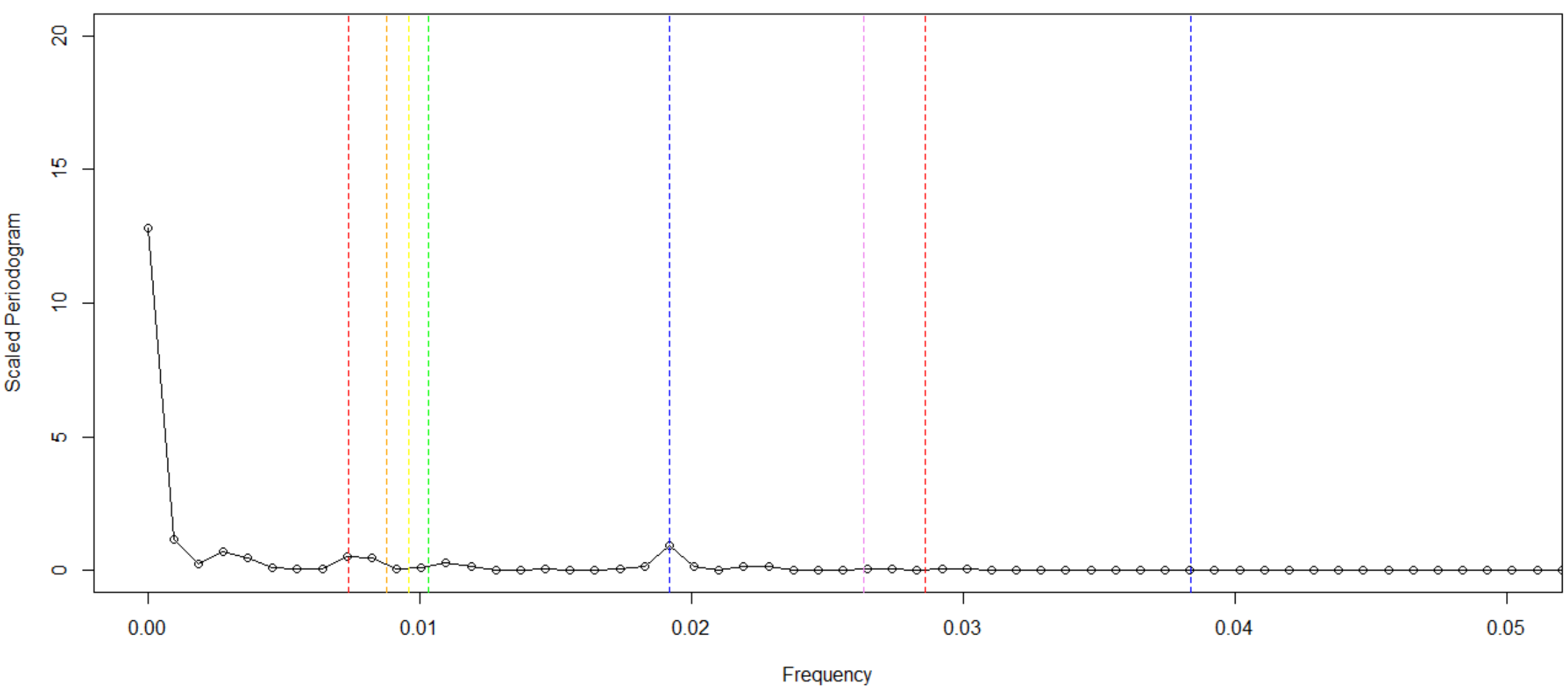

**Figure 11:** The post-vaccine introduction NYC measles rate time series periodogram. Frequencies are indicated from left to right for 136-weeks in red, 114-weeks in orange, 2-year (104-weeks) in yellow, 97-weeks in green, annual (52-weeks) in blue, 38-weeks in violet, 35-weeks in red, and annual second harmonic (26-weeks) in blue.

Of the 14 tested candidate PC components, only two were found significant using VBPBB. The seven remaining PC components that were significant pre-vaccine introduction were no longer significant. No PC components were found significant using PBB.

The 52-week component was found to be significant using VBPBB and a KZFT argument of m=331. It had an estimated signal to noise ratio of 0.10 and provided an estimated 95% CI for the amplitude of the component from (0.44, 2.87). Again, for comparison, PBB could not identify this component as significant. Figure 12 shows several cycles of the 52-week component, the 95% CI band from VBPBB in blue (which excludes a flat line of insignificance) and the KZFT band-pass filtered PC component in green that was used for bootstrapping with VBPBB. The figure also shows the 95% CI band from PBB in red and the original data in yellow used for bootstrapping with PBB.

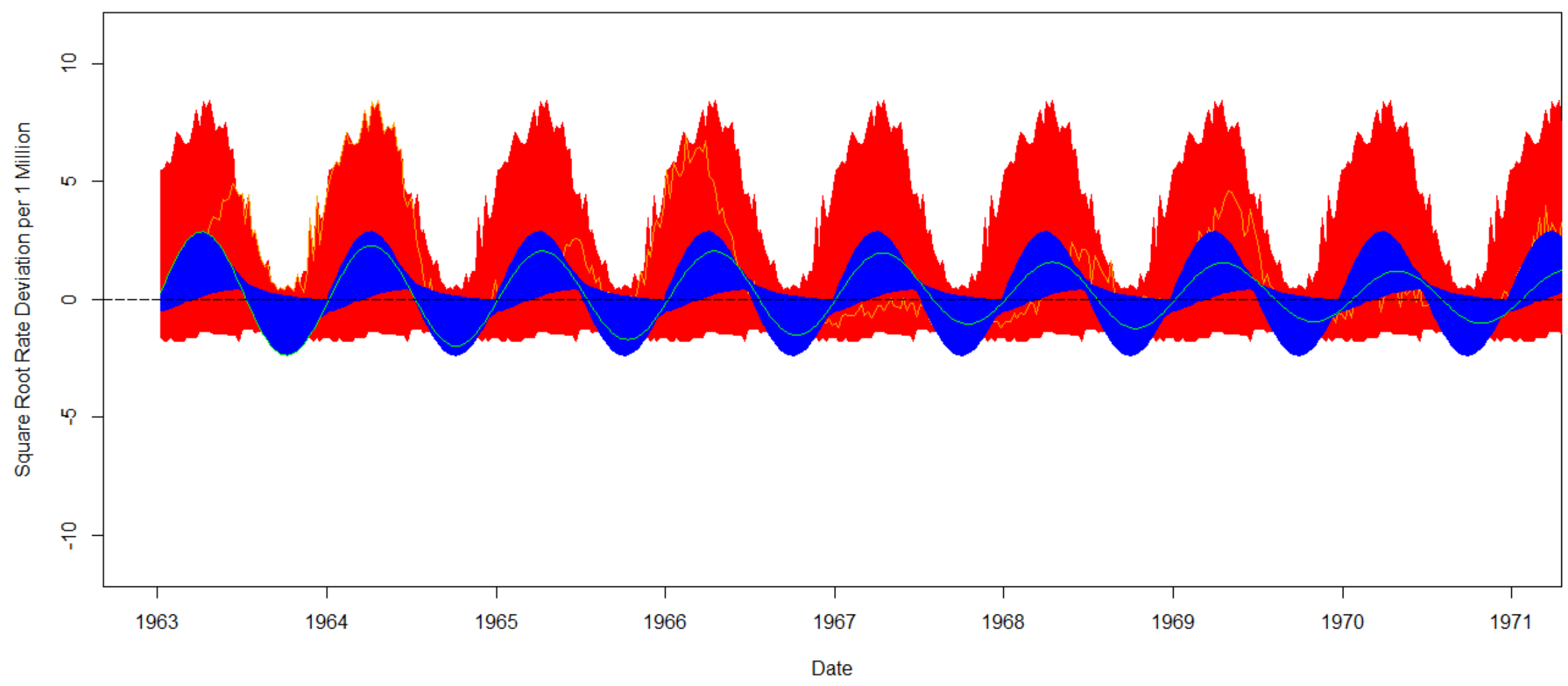

**Figure 12:** Several cycles of the 52-week component of post-vaccine introduction NYC measles rate. The 95% CI band from VBPBB is in blue and the KZFT band-pass filtered PC component is in green. The 95% CI band from PBB is in red and the original data is in yellow.

The 26-week component was found to be significant using VBPBB and a KZFT argument of m=361. It had an estimated signal to noise ratio of 0.004 and provided an estimated 95% CI for the amplitude of the component from (0.02, 0.25). Again, for comparison, PBB could not identify this component as significant. Figure 13 shows several cycles of the 26-week component, the 95% CI band from VBPBB in blue (which excludes a flat line of insignificance) and the KZFT band-pass filtered PC component in green that was used for bootstrapping with VBPBB. The figure also shows the 95% CI band from PBB in red and the original data in yellow used for bootstrapping with PBB.

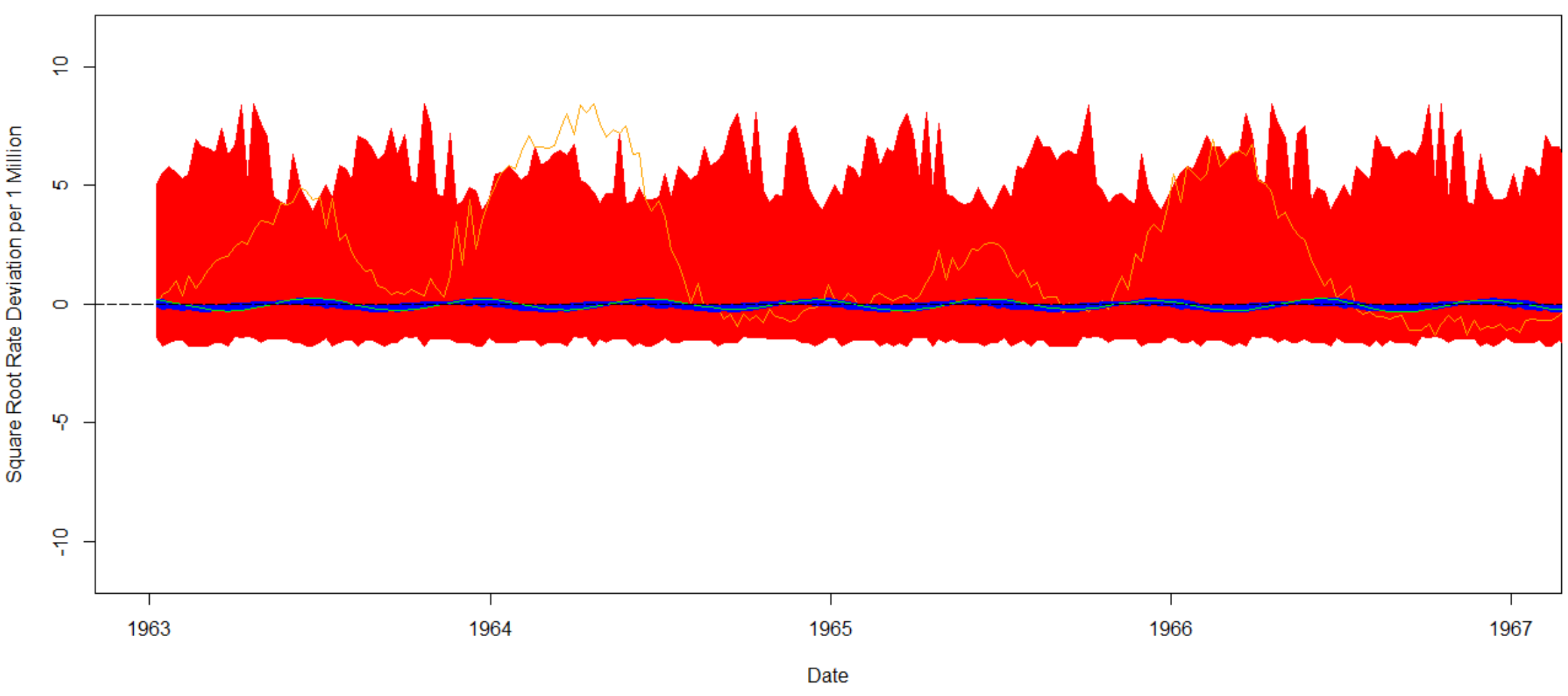

**Figure 13:** Several cycles of the 26-week component of post-vaccine introduction NYC measles rate. The 95% CI band from VBPBB is in blue and the KZFT band-pass filtered PC component is in green. The 95% CI band from PBB is in red and the original data is in yellow.

## Discussion

In this study we investigated the significant PC components of pre- and post-vaccine introduction NYC measles rates using a novel bootstrap resampling method (VBPBB) designed for analysis of time series with periodic and multiple periodic components. Using 93 years of NYC measles time series data from 1891-1984 compiled by Hemple and Earn (2015), we found significant differences between pre- and post-vaccine rate patterns. After splitting the analysis between two datasets, we identified nine significant periodically correlated components in the pre-vaccine NYC measles rate. Alternative methods to the VBPBB could only identify one significant component. One of the strongest components was the annual (52-week) component. The mean square root rate throughout the 1891-1963 pre-vaccine period is approximately seven. The deviation from the annual component in square root pre-vaccine NYC measle rate ranged from (2.82, 5.05). What this translates to is a deviation from the mean case rate of approximately 84 to 140 cases per 1 million population per week. So, throughout the annual cycle, the peak is approximately 84 to 140 cases per 1 million population per week higher than the trough.

In the post-vaccine introduction period from 1963-1984, after testing all of the same potential PC components, we identified two significant PC components in the post-vaccine introduction NYC measles rate. Alternative methods to the VBPBB could not identify any significant components. Again, the strongest component was the annual (52-week) component. However, the strength of this component was greatly reduced compared to the pre-vaccine period. In fact, it appeared all previously significant components were suppressed or muted as the proportion of the NYC population vaccinated increased. The other PC components appear to be suppressed to the point of insignificance. As an example, the mean square root rate throughout the 1963-1984 post-vaccine period is approximately 1.79. The deviation from the annual component in square root post-vaccine NYC measles rate ranged from (0.44, 2.87). What this translates to is a deviation from the mean case rate of approximately 3 to 21.7 cases per 1 million population per week. So, throughout the annual cycle, the peak is approximately 3 to 21.7 cases per 1 million population per week higher than the trough. Recall, pre-vaccine the annual cycle peak is approximately 84 to 140 cases per 1 million population per week higher than the trough.

This is strong evidence that the post-vaccine introduction period suppressed and eliminated many of the significant PC components of measles rates. In the post-vaccine introduction period, as vaccination rates increased, many of the measles rate PC components exhibited greater suppression. This also provides insight into potential PC components which might re-emerge if vaccination rates continue to fall. This can benefit preparation, planning, and timing of potential interventions as previously suppressed components start to re-emerge. These methods could also be repeated for different infectious diseases and provide insight into their characteristic patterns.

Despite the strengths of this analysis, there are limitations. These results are population-based rather than case-based so the conclusions may not apply to individuals or subsets of the population. Additionally, this analysis should be considered preliminary. We recommend a more detailed and thorough analysis to better characterize and confirm the results. This type of analysis would benefit from a greater time span of data including up to the current year. Also, a more detailed pattern may be uncovered using finer data such as daily records. This analysis could benefit from being repeated in different locations. Lessons learned from NYC may serve in preparation to better combat measles and improve public health responses elsewhere. Furthermore, adding a spatial component to the temporal analysis may produce interesting results.

This work provides evidence that as vaccine suppression diminishes, the previously suppressed components may re-emerge. However, this work can serve as a template to translate the timing and severity of measles rate patterns into actionable plans for prevention, preparation, outbreak detection, public health response, resource allocation, and management.